\documentclass{article}

\PassOptionsToPackage{numbers, compress}{natbib}
\usepackage[preprint]{neurips_2019}

\usepackage[utf8]{inputenc} 
\usepackage[T1]{fontenc}    
\usepackage{hyperref}       
\usepackage{url}            
\usepackage{booktabs}       
\usepackage{amsfonts}       
\usepackage{nicefrac}       
\usepackage{microtype}      
\usepackage{amsmath}
\usepackage{graphicx}
\usepackage[table]{xcolor}
\usepackage{enumitem}
\usepackage{pifont}

\title{Silly rules improve the capacity of agents to learn stable enforcement and compliance behaviors}

 \author{%
    Raphael Köster \\
   DeepMind\\
   rkoster@google.com \\
   \And
   Dylan Hadfield-Menell \\
   Department of Electrical Engineering and Computer Science,\\
   University of California Berkeley \\
   Center for Human-Compatible AI\\
    dhm@eecs.berkeley.edu \\
    \texttt{} \\
 \AND
    Gillian K. Hadfield \\
    Schwartz Reisman Institute for Technology and Society,\\
    University of Toronto\\
    Vector Institute\\
    Center for Human-Compatible AI\\
    OpenAI\\
    g.hadfield@utoronto.ca \\
    \texttt{} \\
    \And
    Joel Z. Leibo \\
    DeepMind \\
    jzl@google.com \\
 }

\begin{document}

\maketitle

\begin{abstract}
How can societies learn to enforce and comply with social norms? Here we investigate the learning dynamics and emergence of compliance and enforcement of social norms in a foraging game, implemented in a multi-agent reinforcement learning setting. In this spatiotemporally extended game, individuals are incentivized to implement complex berry-foraging policies and punish transgressions against social taboos covering specific berry types. We show that agents benefit when eating poisonous berries is taboo, meaning the behavior is punished by other agents, as this helps overcome a credit-assignment problem in discovering delayed health effects. Critically, however, we also show that introducing an additional taboo, which results in punishment for eating a harmless berry, improves the rate and stability with which agents learn to punish taboo violations and comply with taboos. Counterintuitively, our results show that an arbitrary taboo (a "silly rule") can enhance social learning dynamics and achieve better outcomes in the middle stages of learning. We discuss the results in the context of studying normativity as a group-level emergent phenomenon. 
\end{abstract}

\section{Introduction}

    Third-party punishment \cite{fehr2004third} of behaviors marked as punishable by a group plays a critical role in the emergence and sustainability of the \emph{ultrasociality} \citep{richerson1998ultrasociality, herrmann2007humans} that characterizes human societies \citep{boyd1992punishment,boyd2010coordinated}. Although non-human primates also engage in cooperative behavior, they do so on a much more limited basis, rooted in reciprocity, and do not display the \emph{normative moralization} \citep{fessler2003meat} of behaviors and third-party punishment that characterize human social orders \citep{riedl2012chimps,tomasello2013review}.  Understanding the nature and dynamics of human normativity---which we define as the systems by which behaviors come to be marked (and possibly unmarked) by a group as punishable and how third-party punishment is coordinated and stabilized---is thus essential for understanding distinctively human intelligence.  Indeed, third-party punishment, which may take forms ranging from mild disapproval or criticism to expulsion from a group or physical violence \citep{wiessner2005norm, boehm2012moral}, may be essential for many aspects of human intelligence, from small-scale kin-based cooperation \citep{mathew2013small} to managing local natural resources \cite{ostrom1990governing}, and even the emergence of language \citep{boyd2015language}.

  Many if not most human norms are functional. Rules that punish non-cooperative behavior, for example, support cooperation.  But an intriguing feature of human normativity is that many social norms concern behaviors that have no direct impact on material well-being.  Examples include rules about what color clothing one wears to a funeral \citep{tomasello2013review} or whether one uses one's left or right hand in particular tasks \citep{fessler2003meat}. Such apparently pointless rules are ubiquitous, often acquiring great social meaning despite the absence of functionality. Hadfield-Menell et al. (2019) call these norms ``silly rules'' and distinguish them from ``important rules,'' such as rules that govern resource sharing or prohibit harmful conduct, that directly impact welfare \cite{hadfield2019legible}.  They demonstrate in a computational experiment that groups with sufficiently cheap silly rules are better able to maintain population in the face of shocks to beliefs about a group's capacity to enforce its important rules. Their account stresses the information benefits of including silly rules in a rule regime: a higher density rule environment provides agents with more opportunities for agents to observe the state of the group, i.e. how effectively it is enforcing its rules (something that may change over time with, for example, changes in group membership). This improves the agents' ability to make optimal decisions about whether to continue to participate in a group or to retreat to a safer, but possibly less valuable, (non-group) setting. Silly rules may thus emerge as a feature of a human normative system because of their contribution to the robustness of groups.
  
  Building on a concept introduced in robotics by \cite{Dragan:2013:LPR:2447556.2447672} to capture the idea that some human actions will be more expressive of human intent---easier to read by the robot--- Hadfield-Menell et al. (2019) suggest that silly rules may increase the \emph{legibility} of a normative system.  Participants are better able to read the state of the group's normative system when they have silly rules available to provide more opportunities to observe normative conduct. That high legibility is a desirable property for a scheme of social norms was also proposed by Axelrod and Hamilton (1984), who emphasized the benefits of "clarity" in cooperative strategies \cite{axelrod}. 
    
    In this paper we introduce the concept of legible normativity to the study of multiagent reinforcement learning. This work is part of a research program that ultimately aims to develop models capable of capturing distinctive features of human intelligence such as the origin of institutions \citep{leibo2019autocurricula}. In particular, we model norm enforcement and compliance behaviors in a multiagent reinforcement learning environment by drawing on a framework developed by \cite{hadfield2012law}. In this framework, a norm is defined as the content of a \emph{normative social order} \citep{hadfield2014microfoundations}. This normative social order is characterized by a binary normative classification scheme---which labels all actions as either punishable or not punishable (violating a norm)---and an enforcement scheme---which penalizes agents that take punishable actions. We consider in particular the enforcement scheme on which, throughout human history, human groups have primarily relied: third-party punishment delivered voluntarily by individual agents. Importantly, the classification scheme is provided by the environment and does not have to be learned by the agents. This implementation is akin to a top-down norm \citep{morris2019norm}, in contrast to norms emerging bottom-up \citep{brooks2011modeling,mahmoud2016cooperation, ajmeri2018robust, morales2013automated}. Historically, systems with institutionalized laws but distributed enforcement have existed. For example, in medieval Iceland the ``lawspeaker'' declared the content of the rules of the society. The enforcement of the rules however was private and rested on the voluntary participation of individuals. \cite{hadfield2013law}.
    
    By providing the environment with the normative system in our experiment, we can then model the aggregate-level phenomenon of a system of rules by looking at the behaviors chosen by individual self-interested, reward-maximizing agents in a group: the actions they choose (with resulting labels) and the punishments they deliver. With this microfoundational account, we do not need to implement any assumptions about agents' cognitive capacity to understand the concept of a 'rule'; nor do we need to impose any particular psychological orientation or preferences for rule-compliance. Compliance with rules simply means that agents in a group avoid behaviors that the group's classification scheme deems punishable.  This occurs only if the group successfully implements its enforcement scheme. This framework thus focuses attention on the fundamental challenge of coordinating effective group punishment: are agents capable of and incentivized to engage in third-party punishment, organized by the group's shared classification scheme?
    
    We consider a simple foraging environment in which a group of agents can individually benefit from compliance with an important rule, specifically a norm against eating a ``poisonous berry''.  We assume that agents do not know which type of berry is poisonous and they have to learn the color of the poisonous berry type. The impact of the poisonous food shows up only at a point many timesteps after its consumption, at which time it reduces the agent's capacity to absorb nutrients. Thus reinforcement learning agents in this environment face a tremendous  credit-assignment problem: attributing the negative effects they experience to a particular berry they ate (among many) several steps earlier. This is a context in which a social rule may be functional: a taboo against eating the berry---enforced by third-party punishment---can improve the payoffs for individual group member. Such a rule may relieve individual agents of the need to rely exclusively on discovering the long-delayed effect for themselves. Effectively, we assume the norm contains the information about the causal relationship and individual agents do not need to discover or understand the causal relationship; they just need to follow the rule, and have an incentive to do so because they will be punished otherwise.
    
Similarly, long-term  credit assignment difficulties abound in the literature on the evolution of culture \citep{giuliano2019understanding}. For example, among some Latin American groups, there is a norm of adding a handful of ash to the cooking pot when preparing maize. Although participants in this practice generally cannot explain why they do this, other than that it is their tradition, adding ash (or limestone) to the cooking water increases the body's ability to absorb niacin. Niacin helps prevent a disease known as pellagra, which over time causes diarrhea, dermatitis, and dementia---and can be fatal. Food taboos appear universal and, by their variability, appear to include both functional taboos that aid nutrition and arbitrary taboos that appear unrelated to nutritional health \citep{meyer2009food}. For a learner, it would be hard to distinguish which nutritional rules are important or arbitrary, given the complexity of the credit assignment problem. Indeed, it has been suggested that cultural evolution can accumulate improving technology without being mirrored by an increased causal understanding of individuals \cite{derex2019causal}.
    
In our study, the environment provides the shared classification scheme: a rule set $R$ deems actions as $R$-wrongful, marking agents that transgress by breaking a food taboo and consuming a forbidden type of berry. $R$ can contain taboos of two kinds: taboos that are directly functional (i.e.~those against poisonous berries), and taboos that are arbitrary (i.e.~those against non-poisonous berries). The enforcement of and compliance with $R$ has to be achieved by a population of simultaneously learning and interacting agents. Agents have to learn to recognize and understand the classification scheme $R$ and learn how to punish transgressions. They will only learn to comply with the rules if the group has learned to punish reliably.
    
Our core question is about the learnability and emergence of punishment and compliance behaviors in connection with properties of the classification scheme $R$. Intuitively, one might hypothesize that a simpler and more functional rule-set $R$---consisting only of important rules---would make the behaviors of punishing and compliance easier and faster to learn. Our results are thus surprising: we show that a rule-set $R$ enriched with silly rules---taboos on eating harmless berries---can in certain circumstances improve agents' capacity to learn the behaviors that generate a normative social order.

    \subsection{Studying norms in a temporospatially extended environment}

    Typically, computational simulations of populations and cultural development utilize an abstracted or idealized space \cite{boyd2003evolution, brooks2011modeling,mahmoud2016cooperation, ajmeri2018robust}. For example, in \cite{hadfield2019legible, hadfield2012law}, the action-space includes ``punishing'' as an atomic action that the agent can choose, and encounters between agents happen in an idealized way. Similarly, an agent directly ``perceives'' a rule violation in an idealized way. In this abstracted setup, it is clear that there is a strong connection between silly and important rules.
    
    Here we explore the role of different social norms in a setting with more fine-grained temporal and spatial substructure. This multi-agent reinforcement learning approach has been successfully used to study intertemporal (sequential) social dilemmas  \cite{kleiman2016coordinate, leibo2017multi, lerer2017maintaining, peysakhovich2017consequentialist, perolat2017multi, hughes2018inequity, foerster2018learning, peysakhovich2018prosocial, wang2019evolving}. Agents inhabit a 2-D grid-world in which they and other objects are located at coordinates in space. The atomic actions in an agent's action-space are moving up, down, left, right, rotating left and right and using a ``punishing beam'' (which allows players to remove rewards from other players, akin to third-party punishment). An agent perceives raw pixels. How these pixels relate to other agents or their actions must be learned. The behavior of the agent is driven by its learning to maximize the expected value of all future rewards it will obtain from its environment (e.g. by collecting berries). This learning over time is accomplished by incremental adjustment of neural network weights. This forms distributed neural representations that produce reward-maximizing behavior in response to visual input of the current situation. Agents learn continuously while being exposed to episode after episode, inhabiting the same environment with a population of other agents who learn simultaneously with them. In order to do this effectively, agents need to correctly assign credit to current stimuli and actions based on subsequent rewards they receive. This creates a rich dynamic in which every part of a behavior has to be learned, and strategic decisions have to be \emph{implemented} via a behavioral policy. Both the cognitive challenge of correct credit assignment as well as performing complex action sequences are difficult and the dynamics of how norms are learned and implemented are endogenous to the multi-agent learning model. This leads to a number of important differences from more abstracted simulations like matrix games and affords rich opportunities to study the emergence and importance of social norms in ways that could not otherwise be approached:
    \begin{enumerate}
        \item Punishing other agents' behavior or observing a rule violation are complex sequences of actions or stimuli that look different each time they are performed or encountered.
        \item As everything has to be learned, we can expect a temporal dependency and hierarchy among learned behaviors. For example for agents to learn to avoid a social taboo, agents will first need to learn how to effectively apply punishing, which they can only learn after they have learned to visually parse the world and accurately move through it.  
        \item As agents are driven by maximizing total reward, whether or not an agent engages in social punishing depends on the opportunity cost of the action sequence, the agent's skill in implementing it, and the reward gained by punishing the other agent's transgression.
        \item As the social dynamics of punishing silly or important social taboos need to be implemented in similar ways and are learned in neural networks from scratch, they afford the opportunity for generalization during learning.
        \item As social punishing of silly or important rules is implemented in the same way, a confusion between the two can arise. Similarly, punishing might be misdirected at agents that did not break a social taboo. These costly false-positive incidents provide a counterweight to the development of an indiscriminate social punishing dynamic. Importantly, mistakes in behavior are not modelled as noise that is injected externally, but are emergent from the learning dynamics and the inherent difficulty of implementing an effective behavior policy.
        \item Legibility of social norms and their enforcement does not just depend on the number of taboos, but interacts with the inherent difficulty of having to learn perception, credit assignment, and behavior. For example the ability to learn from norms might be related to the agents' field of vision, their ability to process the visual information, how often they observe rules being broken and their ability to remember and understand the rules. 
    \end{enumerate}

  We hypothesize that in the reinforcement-learning context silly rules can be beneficial for a group because they provide more data---or legibility---about punishing and norms in general. Silly rules can support the learning dynamics of the group by providing more opportunities for agents to learn the fundamental skill of engaging in third-party punishment of rules that support welfare for individuals and groups.

\section{Methods}

\subsection{Multi-Agent Reinforcement Learning}

We consider multi-agent reinforcement learning in partially-observable general-sum Markov games \cite{shapley1953stochastic, Littman94markovgames}. In each game state, agents take actions based on a partial observation of the state space and receive an individual reward. Agents must learn through experience an appropriate behavior policy while interacting with one another. We formalize this as follows: an $N$-player partially observable Markov game $\mathcal{M}$ defined on a finite set of states $\mathcal{S}$.
The observation function $\mathcal{O} : \mathcal{S} \times \{1, . . . , N\} \rightarrow \mathbb{R}^d$, specifies each player's $d$-dimensional view on the state space.

In each state, each player $i$ is allowed to take an action from its own set $\mathcal{A}^i$.

Following their joint action 
$(a^1, . . . , a^N) \in \mathcal{A}^1 \times \! . . . \! \times \mathcal{A}^N$,
the state changes obeys
the stochastic transition function 

$\mathcal{T} : \mathcal{S} \times \mathcal{A}^1 \times \! . . . \! \times \mathcal{A}^N \rightarrow \Delta(\mathcal{S})$, where $\Delta(\mathcal{S})$ denotes the set of discrete probability distributions over $\mathcal{S}$, and every player receives an individual reward defined as 

$r^i: \mathcal{S} \times \mathcal{A}^1 \times . . .  \times \mathcal{A}^N \rightarrow \mathbb{R}$
for player $i$. 
Finally, let 

$o^i = \{\mathcal{O}(s, i)\}_{s \in \mathcal{S}}$ 
be the observation space of player $i$.

Each agent learns, independently through its own experience of the environment, a behavior policy $\pi^i : \mathcal{O}^i \rightarrow \Delta(\mathcal{A}^i)$ (written $\pi(a^i|o^i)$) based on its own observation $o^i = \mathcal{O}(s,i)$ and extrinsic reward $r^i(s,a^1,\dots,a^N)$. Each agent's goal is to maximize a long term $\gamma$-discounted payoff defined as follows:
\begin{equation}
V_{\vec{\pi}}^i(s_0) = \mathbb{E} \left[ \sum \limits_{t=0}^{\infty} \gamma^t r^i(s_t, \vec{a}_t) | \vec{a}_t \sim \vec{\pi}_t, s_{t+1} \sim \mathcal{T}(s_t, \vec{a}_t) \right] \, .
\end{equation}

\subsection{Experiment and conditions}

We consider a foraging task implemented as a partially observable Markov game on a 2D grid (see Fig. 1). Agents gain reward by collecting berries that stochastically respawn. The respawn probabilities are high, so there is little competition for resources. Moving onto the coordinates of a berry, agents earn a reward of 4 points. Each berry type is consistently mapped to a color (24 different types in the standard setting). One berry type is "poisonous". There is no other signal of which berry type is poisonous that is observable to an agent at the time of consumption, except the color that remains consistent for all episodes.  If collected by a player, this player is "poisoned" after a delay of a fixed number of timesteps (100 timesteps in the standard setting). Poisoning reduces a player's ability to absorb nutrition: after poisoning sets in each subsequent berry the player collects yields a reward of 1 instead of 4. Besides moving, agents have in their behavioral repertoire the ability to apply a "punishing beam". If successfully targeted at another player, the user of the beam loses a reward of 20 (the cost of punishing, in addition to the opportunity cost of time spent aiming and firing the beam instead of collecting berries) and the punished player loses a reward of 35 \footnote{Video of example episode:  \href{https://youtu.be/Xn2eTSX-4GU}{https://youtu.be/Xn2eTSX-4GU}. Consumption of taboo berry and subsequent punishment at 23-25 seconds. Note that agents see a lower resolution version of the environment in which each entity is represented by a single pixel.}. 

Each instance of the training regime is initialized in one of 3 different conditions. This is a between-subjects design: each agent population only experiences one of these 3 conditions. The conditions differ in the content of the classification scheme $R$ that marks agents if they have broken a taboo. Let $R: \{\text{berry types}\} \rightarrow \{0,1\}$ be a function that classifies berry eating events as either right (1) or wrong (0). We adopt the notation $R\{a,b\}$ to indicate an experimental condition where eating berry types $a$ and $b$ are marked wrongful while all other berries may be rightfully consumed. The first entry refers to the poisonous berry type, the second refers to a non-poisonous berry type, leading to three different cases we consider: $R\{\}$, $R\{\text{poisonous}\}$, $R\{\text{poisonous, nonpoisonous}\}$.

In the "norm-free" condition, $R\{\}$, there are no additional mechanics to the game beyond what is described above. Agents have to learn which berry is poisonous without any additional information. 

In the "Important rule" condition, $R\{\text{poisonous}\}$, we introduce a group rule against eating the poisonous berry type. In this condition, a player that eats a poison berry is ``marked'': from the perspective of other agents in the environment, the marked player changes color. This color change is not visible to the marked player. This color change implements the idea that other agents evaluate the consumption behavior of the agent that has chosen to eat a 'taboo' food. This marking then interacts with the punishing capacity of other agents. If a "marked" player is successfully targeted by another player with a punishing beam, the punishing player gets a reward of 35---effectively transferring reward from the marked player to the punishing player, for a net payoff to the punishing player of 15 points (note that when considering the sum of rewards of the whole group, a successful punishment results net-loss for the group of 20 points because of the cost of using the punishment beam). Aiming punishment at a non-marked player is costly to both as in the $R\{\}$ condition. Once punished, the marking disappears.

In the third condition, $R\{\text{poisonous, nonpoisonous}\}$, we augment the important rule with an additional silly rule, or arbitrary taboo.  In this condition, players become marked not only if they consume the poisonous berry but also if they consume another designated, but harmless, berry. As in the $R\{\text{poisonous}\}$ condition, successful punishing of an agent that has violated the silly rule by consuming the designated harmless berry earns the punishing agent a net of 15 points and costs the transgressing agent 35 points.  Thus, from the perspective of the agents, the 'important' and 'silly' rules are isomorphic if they have not integrated knowledge of the actual poisoning dynamic.  

Note that in these settings classification scheme of $R$ is implemented by the environment.  We have not modeled the emergence of the rules in themselves. Agents are incentivised to learn policies that implement the behaviors of collecting berries, delivering third-party punishment, and avoiding taboo berries that create a risk of punishment.

  \begin{figure}[ht]
  \begin{center}
  \includegraphics[width=\textwidth]{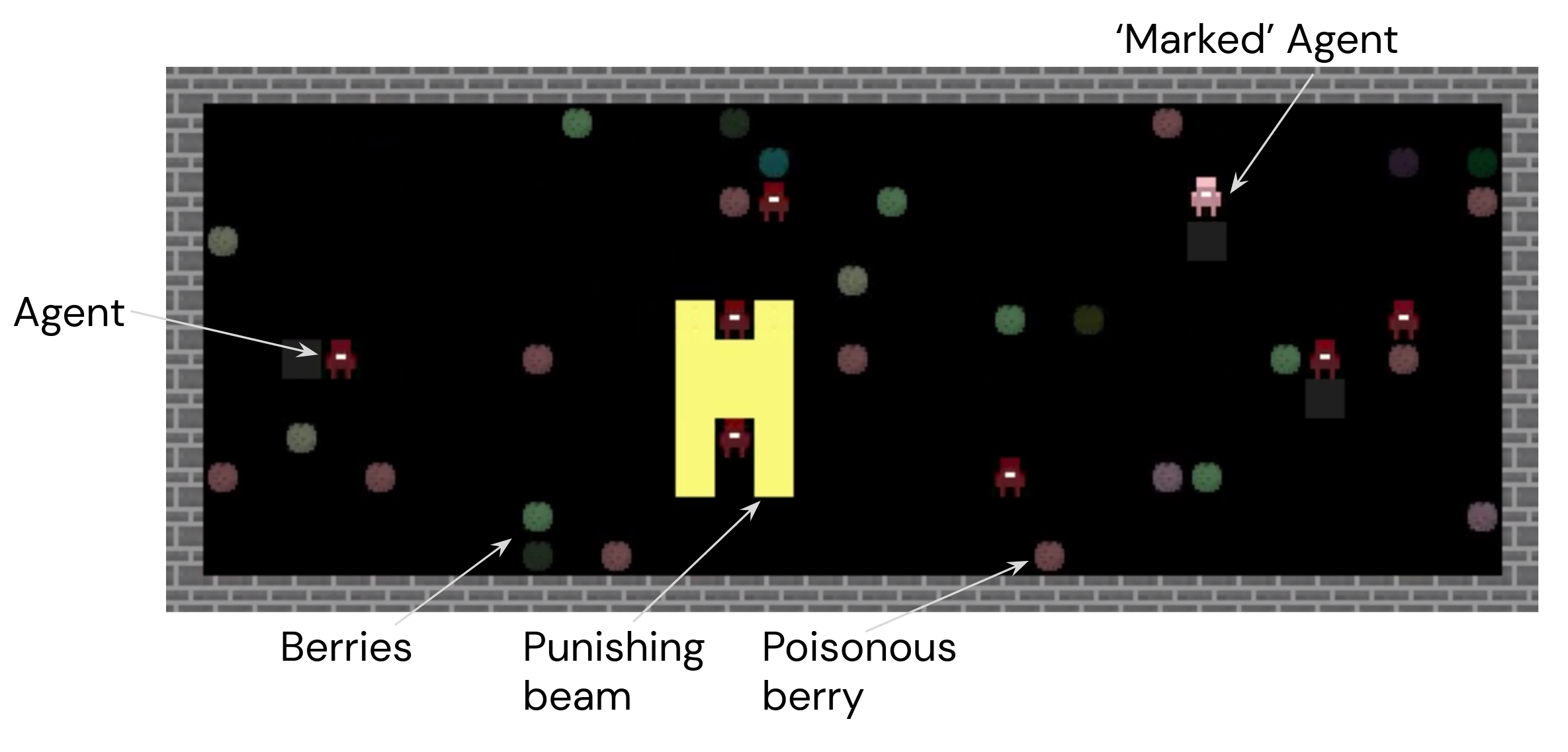}
  \end{center}
  \caption{Depiction of the environment. The agents inhabit a grid world. Agents earn reward for eating berries, which regrow continuously. One type of berry is poisonous and if collected by an agent, it diminishes the agent's ability to gather rewards from other berries, after a delay period. If an agent eats one of the poisonous berries in the R\{poisonous\} condition, the agent immediately gets "marked" and appears in a different color to the other agents. In the R\{poisonous, nonpoisonous\} condition, one additional, non-poisonous, berry also triggers an agents' marking. Agents are able to punish each other using a "punishing beam", causing a loss to themselves and a large loss to the punished agent. If a "marked" agent is punished, the punishing agent receives a large reward.}
\end{figure}

    \subsection{Agent architecture and training method}
    
     Each instance of the training regime contained a population of 12 learners. The environment is a gridworld of size $33 \times 12$ pixels and agents observe a $15 \times 15$ pixels RGB window, centered on their current location (note that the depictions in this paper are higher resolution for display purposes). On each episode, a subset of learners was drawn without replacement to play in the current episode. The standard setting we consider in detail contained 8 players in each episode. Each episode lasted for 1000 steps. For each timestep $s$, each learner $i$ in the population produced a policy $\pi^i$ and an estimate of the value $V_{\vec{\pi}}^i(s)$ with a neural network, implemented on a GPU. This neural network was trained by receiving importance-weighted policy updates \cite{espeholt2018impala} sampled from a queue of trajectories. These trajectories were created by 64 simultaneous environments on CPUs that play the game (with 8 players in the standard setting, which used policies sampled uniformly from the population of learners without replacement). The learners received truncated sequences of 100 steps of trajectories in batches of 16.
    
    The neural network's architecture consisted of a visual encoder (2D-convolutional neural net with 6 channels, with kernel size and stride size 1) followed by a 2-layer fully connected MLP with 64 RELU-neurons in each layer, an LSTM (128 units) and finally linear policy and value heads, outputting the value of the current state and a probability over actions to be chosen. We used a discount-factor of 0.99, the learning rate was 0.0004, and the weight of entropy regularisation of the policy logits was 0.003. We used the RMS-prop optimiser (learning rate=0.0004, epsilon=1e-5, momentum=0.0, decay=0.99). The agent also minimized a CPC loss \cite{oord2018representation} in the manner of an auxiliary objective \cite{jaderberg2016reinforcement}.

\section{Results}

As displayed in Fig. 2, we examine group-level metrics about agent-populations over the trajectory of learning. We plot the average trajectory per condition. As visible in Fig. 2A, the first thing agent populations learn is to reduce the amount that unmarked players are punished. Punishing unmarked players is costly to both the punished and the punishing agent, so it is unsurprising that this behavior does not persist long once actions become less random. As can be seen in Fig. 2F, this rapid initial learning increases the collective return. The collective return is the sum of rewards gained by all agents. Note that the suppression of misdirected punishing happens fastest in the $R\{\}$  condition. This is unsurprising, as in this condition there is no direct incentive to punish any other players at all, because there are no taboos that lead to marked players. 

  \begin{figure*}[ht]
  \includegraphics[width=\textwidth]{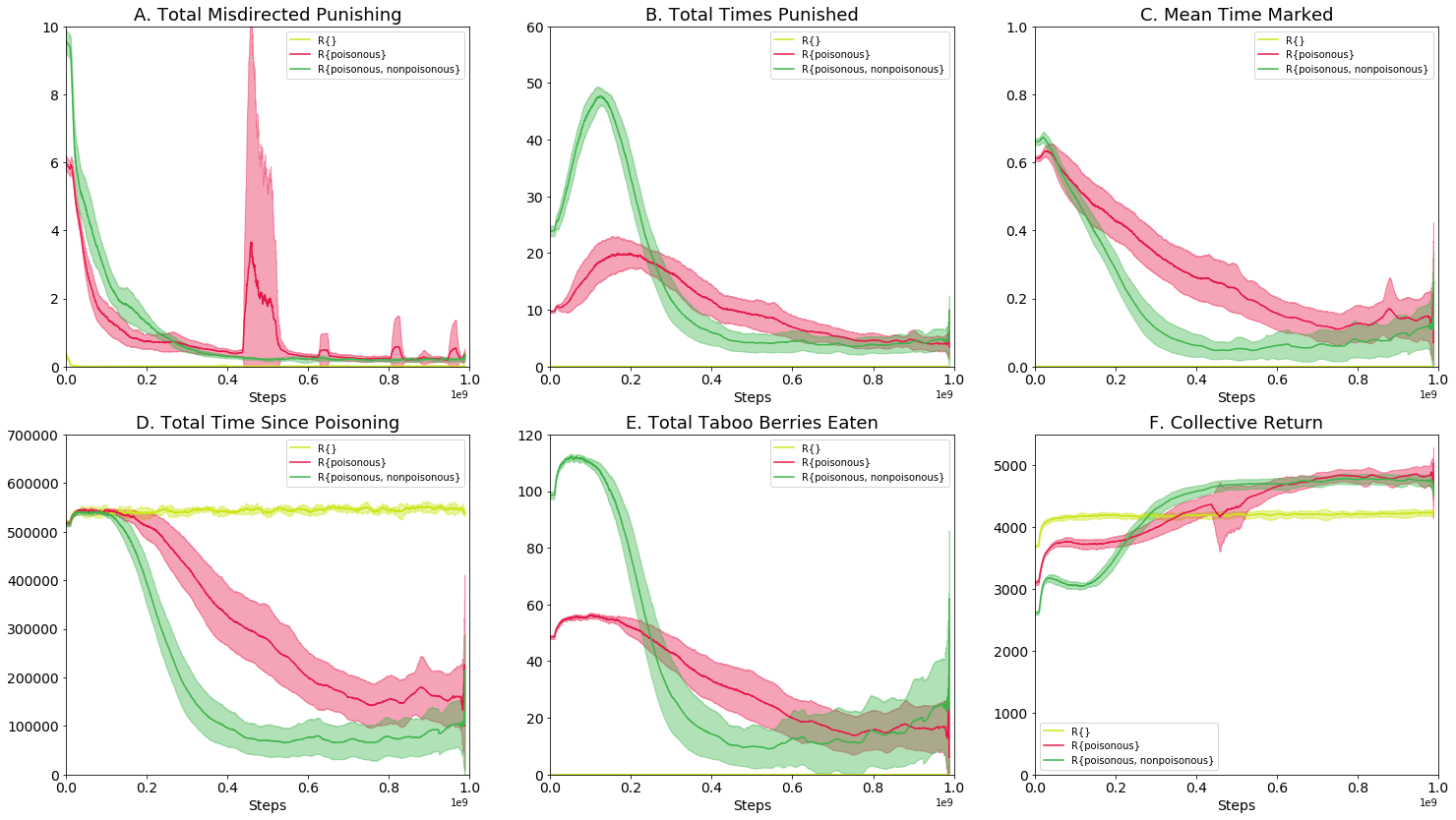}
  \caption{Learning dynamics: We are examining group-level metrics about agent-populations (y-axis) over the trajectory of learning (x-axis in timesteps). We plot the average trajectory per condition (with 99\% confidence interval). 
  A. Number of times unmarked agents are punished (agents that have not broken a taboo). 
  B. Number of times marked agents are punished (agents that have broken a taboo). 
  C. Time spent marked after breaking a taboo.
  D. Time spent since eating poisonous berry. 
  E. The amount of "taboo" berries eaten (poisonous and non-poisonous combined, if available in the condition).
  F. Total sum of reward gained by group (including costs of punishing). In total, we observe a benefit of the \emph{R}\{poisonous, nonpoisonous\} condition in the intermediate stages of learning, driven by an increased ability to avoid poisonous berries. We also see a temporal order to learned behaviors, e.g. an increase of social punishment that then declines together with a degrees of taboo berries eaten.} 
\end{figure*}

The second important learning dynamic is that the number of times `marked players' get successfully punished initially strongly increases before it decreases (Fig. 2B). We interpret the increase as an improvement in the agents' skill at enforcing the social norm, i.e. being increasingly skilled at effectively punishing marked agents. As displayed in Fig. 2C, the amount of time agents spend marked is steadily declining. However, taken by itself, this metric does not differentiate between whether this decline is driven by agents becoming better at avoiding rule violation, or whether agents get better at punishing rule breakers and thereby removing their mark. As can be seen in Fig. 2E, the decline of successful punishments coincides with a decline in the number of taboo berries eaten. This shows that there is a hierarchy in the learned behaviors, as first the social punishing system needs to be successfully implemented before it is possible for agents to learn that they should avoid breaking the social norm. In these two measures (successful punishments and taboo berries eaten) we see the role of the arbitrary taboo (one additional taboo berry) most clearly. Early in learning, it is unsurprising that double the amount of taboo berries leads to a higher amount of taboo berries eaten and subsequent punishing. Interestingly, once these quantities start to decline, they decline more rapidly in the condition with two taboos instead of one and in fact reach a lower level. So, it appears that increased exposure to taboo berries and punishing early leads to more robust learning. This is evident in later stages of learning where agents eat fewer taboo berries in the condition in which there are twice as many. 

As can be seen in Fig. 2D, in terms of avoiding getting poisoned, having two taboos instead of one consistently leads to better results. Additionally, this plot shows that the credit assignment problem of avoiding the poisonous berry without the help of a social punishing mechanism (the yellow line depicts the $R\{\text{}\}$ condition) is so difficult that agent populations do not learn to avoid the poisonous berry. However, the consistent benefit of the additional arbitrary taboo in terms of avoiding the poisonous berries does not in itself translate into a benefit in collective return (Fig. 2F). Collective return sums all rewards gained by all agents. If poisonous berries were avoided by agents just standing still or moving more slowly, the collective return would reveal that agents have not learned to forage successfully. Similarly, collective return factors in the cost of social punishing. If a marked agent is successfully punished, the group loses a total of 20 points. If an unmarked agent is punished, it costs the group 55 points (20 as the cost of punishing and 35 for the punished). This means that in order to achieve a benefit in collective return, the avoidance of poisonous berries has to be so substantial that it surpasses the costs associated with the social punishment scheme. AS shown in Fig. 2F, this is actually the case in the intermediate learning stages. In order to assess the difference between conditions, we divide the learning timecourse into 10 bins and average the collective returns for each instance of agent populations in each bin. We then can use a t-test to compare the $R\{\text{poisonous}\}$  and $R\{\text{poisonous, nonpoisonous}\}$ conditions in each bin. There is a significant benefit of the arbitrary rule condition in the 3rd, 4th and 5ths timebin 
(3rd: t(28)=3.94, p=0.0005, 4th: t(28)=3.26, p=0.003, 6th: t(28)=2.43, p=0.022. The 3rd and 4th timebin remain significant after Bonferonni-correction for 10 multiple comparisons). As visible in Fig. 2D, this benefit of the additional arbitrary taboo is driven by avoiding poisonous berries, but there is another contributing factor. As foreshadowed in Fig. 2A., we observe a benefit of the $R\{\text{poisonous, nonpoisonous}\}$  condition which involves the reduction of misdirected punishing. Fig. 3 displays the trajectories of single training regimes (separate agent populations) that were run in the $R\{\text{poisonous}\}$  and $R\{\text{poisonous, nonpoisonous}\}$  condition. Only in in the $R\{\text{poisonous}\}$  condition it can be observed that some agent populations fall back in a regime of high misdirected punishing. In contrast, in the $R\{\text{poisonous, nonpoisonous}\}$  condition, once high accuracy  of punishing is achieved (i.e. only marked agents are punished), the agents retain this accuracy and do not start to punish indiscriminately. 

  \begin{figure}[ht]
  \begin{center}
  \includegraphics[width=0.8\textwidth]{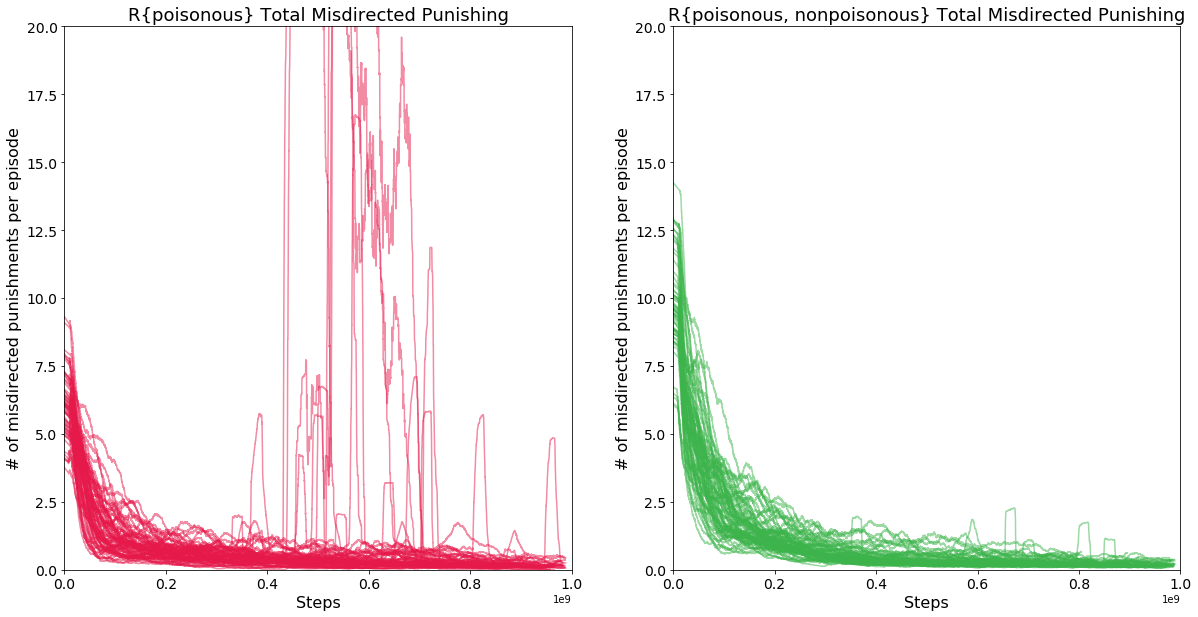}
  \end{center}
  \caption{More robust learning of accurate punishing behavior: Each trajectory is one agent population, plotting the amount of misdirected punishing behavior (punishing unmarked players) over learning. Since the \emph{R}\{poisonous, nonpoisonous\} condition has two taboo berries, it starts out with higher punishing behavior overall, including indiscriminate punishing of unmarked agents. Once high levels of accuracy have been reached, punishing of unmarked players is rare in both conditions. Counterintuitively, this accuracy appears to be more stable in the \emph{R}\{poisonous, nonpoisonous\} condition, and the populations do not relapse into states of high indiscriminate punishing.}
\end{figure}

Finally, we investigated different hypotheses of how exactly arbitrary rules benefit learning. In order to study this question, we ran instances of agent populations with different settings of the environment. We then extracted the average score in the $R\{\text{poisonous}\}$  and \emph{R}\{poisonous, nonpoisonous\}  condition in the 3rd-5ths timebin and averaged the collective return that was obtained (the choice of timebin was based on the results obtained in the default setting, see above, and is orthogonal to the hypothesis tested here). We submitted the results to a between-subjects ANOVA in which each agent population represents a datapoint ("subject") and the factors are "condition", with factor levels: $R\{\text{poisonous}\}$, and \emph{R}\{poisonous, nonpoisonous\} and the setting we vary: number of players, number of berry types or the delay until poison takes effect. The hypothesized effect of a modulation of the relative benefit of the \emph{R}\{poisonous, nonpoisonous\} condition is captured by the interaction effect of the "condition" factor and the factor containing the environment setting (see Fig. 4).

Given that silly rules are a group-level punishing dynamic, we reasoned that the benefit of arbitrary taboos can be influenced by factors underpinning the group level dynamic. In particular, we varied the number of players that inhabit the environment at the same time. We hypothesized that a larger number of players would be associated with a greater benefit of the 'silly rule' condition, because it strengthens the social learning dynamics. For example, as the environment is more densely populated, players observe rule violations more often and have to take fewer steps to punish a marked agent. As predicted, the number of players (factor levels: 6, 7, 8, 9, 10 players) modulates the relative benefit of adding an arbitrary taboo (F(4,58)=3.14, p=0.02). This benefit of arbitrary taboos tends to be larger for higher numbers of players (but is not monotonic).

Second, we reasoned that the benefit or cost of arbitrary taboos have to be weighed against what is gained or lost for the individual's goals, i.e. avoiding poisonous berries. That is, because arbitrary taboos create dead-weight loss due to unnecessary punishing, arbitrary taboos are more likely to be worth the effort if the learning problem posed to individuals is very hard. We used two different methods to make the credit-assignment problem for agents harder. We varied the number of berries and we varied the time delay it took for the poison to take effect. We hypothesized that in the harder conditions (more berries or longer delays), the relative benefit of adding an arbitrary rule would be greater, because the initial learning problem is harder. As predicted, for number of berry types (factor levels: 16, 20, 24 and 28 types of berries, F(3,52)=2.98, p=0.04) and time delay of poisoning (factor levels: 50, 75, 100, 125, 150 timesteps until poison takes effect, F(4,60)=4.36, p=0.004) there was an interaction of condition and environment setting. The benefit of silly rules tends to be larger in the environment settings that lead to a harder credit assignment problem (note however that the relationship is not monotonic). In sum, we show that the environment settings can affect the benefit of arbitrary taboos in different ways, i.e. affecting how easy it is to implement a social norm (having a denser population), or whether it is worth doing so (when facing a harder problem).

  \begin{figure}[ht]
  \begin{center}
  \includegraphics[width=0.6\textwidth]{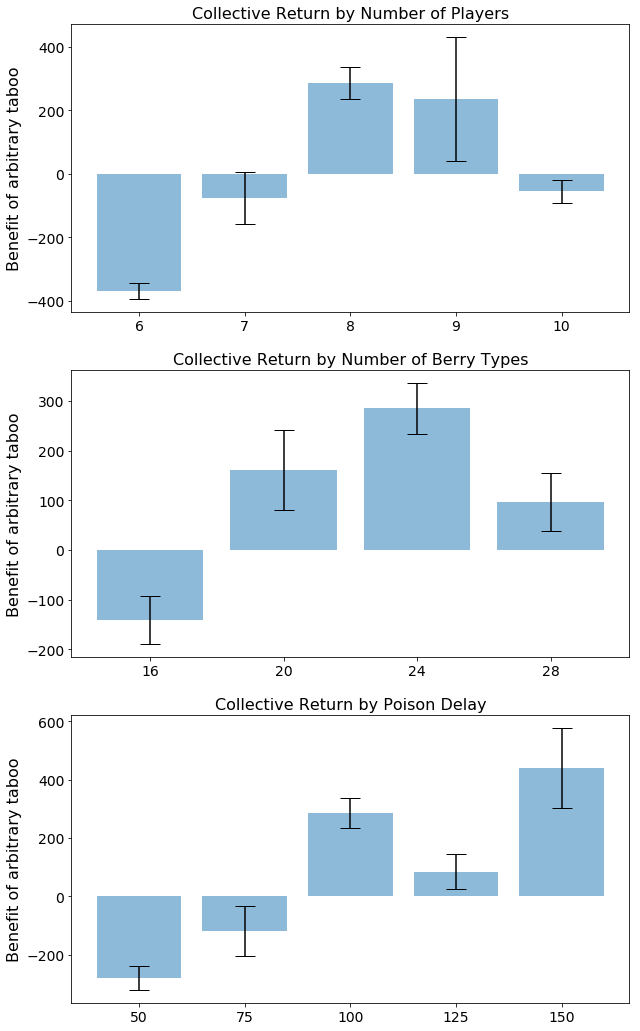}
  \end{center}
  \caption{The benefit of arbitrary taboos is modulated by environmental parameters. We consider the benefit of adding an arbitrary taboo in middle stages of learning by comparing the collective return achieved in the \emph{R}\{poisonous, nonpoisonous\} condition above the \emph{R}\{poisonous\} condition. We plot the means across different environment settings with error bars representing the pooled standard error of the mean. 
  We varied the number of players that inhabit the environment at the same time (top panel). The benefit of arbitrary taboos is highest with 8 players and tends to be larger with a higher number of players, consistent with strengthened social learning dynamics. We varied the number of berry types in the environment (middle panel) and the time delay until the poison takes effect (bottom panel). The settings that lead to a harder credit-assignment problem (more berries and longer time delay) tend to strengthen the relative benefit of an arbitrary taboo.}
\end{figure}
    
\section{Discussion}
  
  We set out to study the enforcement of and compliance with an exogenous normative social order in a setting in which reward-maximizing agents have to learn from scratch and act via complex behavioral policies. We demonstrate that a more complex rule set containing an arbitrary taboo, or silly rule, can lead to faster and more stable learning for reinforcement learning agents, supporting the initial finding in \citep{hadfield2019legible}. This line of research connects to the study of human cultural evolution and social norms that support complex group behaviors such as cooperation. We offer an explanation why arbitrary taboos may appear and are maintained, grounded in the mechanics of learning within a single group. This account is independent of, but not necessarily inconsistent with, existing explanations centered around in-group/out-group classification and group cohesion \cite{meyer2009food}.
  
  While the arbitrary taboo provided a consistent benefit in avoiding poisonous berries, it is worth noting that the benefit of the arbitrary rule on the overall prosperity of the group was only present in the intermediate stages of learning. This could be associated with the dead-weight cost of maintaining a social norm that does not serve a direct material function, or imprecise strategies to avoid poison (i.e. moving more slowly in general). A persistent benefit of the added arbitrary rule appears to be a lower frequency of misdirected punishing. While all agent populations quickly learn to minimize indiscriminate punishing, without the additional arbitrary taboo, populations appear to suffer from lapses back into regimes of more inaccurate punishing. These instabilities are common in deep reinforcement learning setups \cite{sutton2018reinforcement}. Future work could investigate whether these relapses are associated to the better consolidation of accurate policies, driven by the improved training of the condition that contains the arbitrary taboo, or whether there is another role of indiscriminate punishment.
  
  Our findings also echo results from the literature on cultural evolution that suggest larger group sizes can benefit learning and accumulation of culture \cite{henrich2004demography, powell2009late,derex2013experimental}. In our account, this is due to the fact that larger groups, with higher population density, assist agents in learning to participate in the fundamental enforcement scheme. A higher density could benefit learning by providing shorter paths to marked agents and increased number of observations of rule violations and subsequent punishing. Future experiments and analyses could examine the mechanics of this effect in more detail. 
  
  A clear limitation of this work is that we have not shown the emergence of the social norms themselves.  We supplied in the environment the causal relationship between an action---eating a particular berry---and the trigger for social punishing: becoming marked in the view of other agents and generating a reward for an agent who successfully aimed the punishing beam at the transgressor.  The next steps in this work are therefore to attempt to demonstrate the emergence of particular patterns of marking---norms---and the capacity for norms to change in response to changes in the environment or other sources or variation including natural drift. We hypothesize that learning how to follow and maintain social norms can assist agents in adapting to variation in the environment. This social technology of benefiting from norms is closely related to the cultural niche \cite{boyd2011niche} inhabited by humans, and to humanity's intelligence and success. Understanding how this technology emerges in multi-agent settings may play a critical role in understanding the emergence of human-level intelligence.

\small
\bibliography{refs}{}

\begin{thebibliography}{10}

\bibitem{ajmeri2018robust}
Nirav Ajmeri, Hui Guo, Pradeep~K Murukannaiah, and Munindar~P Singh.
\newblock Robust norm emergence by revealing and reasoning about context:
  Socially intelligent agents for enhancing privacy.
\newblock In {\em IJCAI}, pages 28--34, 2018.

\bibitem{axelrod}
Robert~M Axelrod and W~D Hamilton.
\newblock {\em The evolution of cooperation}.
\newblock Basic Books, New York, 1984.

\bibitem{boehm2012moral}
Christopher Boehm.
\newblock {\em Moral origins: The evolution of virtue, altruism, and shame}.
\newblock Soft Skull Press, 2012.

\bibitem{boyd2010coordinated}
Robert Boyd, Herbert Gintis, and Samuel Bowles.
\newblock Coordinated punishment of defectors sustains cooperation and can
  proliferate when rare.
\newblock {\em Science}, 328(5978):617--620, 2010.

\bibitem{boyd2003evolution}
Robert Boyd, Herbert Gintis, Samuel Bowles, and Peter~J Richerson.
\newblock The evolution of altruistic punishment.
\newblock {\em Proceedings of the National Academy of Sciences},
  100(6):3531--3535, 2003.

\bibitem{boyd2015language}
Robert Boyd and Sarah Mathew.
\newblock Third-party monitoring and sanctions aid the evolution of language.
\newblock {\em Evolution and Human Behavior}, 36:475--479, 2015.

\bibitem{boyd1992punishment}
Robert Boyd and Peter~J Richerson.
\newblock Punishment allows the evolution of cooperation (or anything else) in
  sizable groups.
\newblock {\em Ethology and sociobiology}, 13(3):171--195, 1992.

\bibitem{boyd2011niche}
Robert Boyd, Peter~J. Richerson, and Joseph Henrich.
\newblock The cultural niche: Why social learning is essential for human
  adaptation.
\newblock {\em PNAS}, 108(Supp. 2):10918--10925, 2011.

\bibitem{brooks2011modeling}
Logan~Conrad Brooks, Wayne Iba, and Sandip Sen.
\newblock Modeling the emergence and convergence of norms.
\newblock In {\em Twenty-Second International Joint Conference on Artificial
  Intelligence}, 2011.

\bibitem{derex2013experimental}
Maxime Derex, Marie-Pauline Beugin, Bernard Godelle, and Michel Raymond.
\newblock Experimental evidence for the influence of group size on cultural
  complexity.
\newblock {\em Nature}, 503(7476):389, 2013.

\bibitem{derex2019causal}
Maxime Derex, Jean-Fran{\c{c}}ois Bonnefon, Robert Boyd, and Alex Mesoudi.
\newblock Causal understanding is not necessary for the improvement of
  culturally evolving technology.
\newblock {\em Nature human behaviour}, 3(5):446, 2019.

\bibitem{Dragan:2013:LPR:2447556.2447672}
Anca~D. Dragan, Kenton~C.T. Lee, and Siddhartha~S. Srinivasa.
\newblock Legibility and predictability of robot motion.
\newblock In {\em Proceedings of the 8th ACM/IEEE International Conference on
  Human-robot Interaction}, HRI '13, pages 301--308, Piscataway, NJ, USA, 2013.
  IEEE Press.

\bibitem{espeholt2018impala}
Lasse Espeholt, Hubert Soyer, Remi Munos, Karen Simonyan, Volodymir Mnih, Tom
  Ward, Yotam Doron, Vlad Firoiu, Tim Harley, Iain Dunning, Shane Legg, and
  Koray Kavukcuoglu.
\newblock Impala: Scalable distributed deep-rl with importance weighted
  actor-learner architectures, 2018.

\bibitem{fehr2004third}
Ernst Fehr and Urs Fischbacher.
\newblock Third-party punishment and social norms.
\newblock {\em Evolution and human behavior}, 25(2):63--87, 2004.

\bibitem{fessler2003meat}
Daniel~MT Fessler and Carlos~David Navarrete.
\newblock Meat is good to taboo.
\newblock {\em Journal of Cognition and Culture}, 3(1):1--40, 2003.

\bibitem{foerster2018learning}
Jakob Foerster, Richard~Y Chen, Maruan Al-Shedivat, Shimon Whiteson, Pieter
  Abbeel, and Igor Mordatch.
\newblock Learning with opponent-learning awareness.
\newblock In {\em Proceedings of the 17th International Conference on
  Autonomous Agents and MultiAgent Systems}, pages 122--130. International
  Foundation for Autonomous Agents and Multiagent Systems, 2018.

\bibitem{giuliano2019understanding}
Paola Giuliano and Nathan Nunn.
\newblock Understanding cultural persistence and change.
\newblock {\em National Bureau of Economic Research}, 2019.

\bibitem{hadfield2012law}
Gillian~K Hadfield and Barry~R Weingast.
\newblock What is law? a coordination model of the characteristics of legal
  order.
\newblock {\em Journal of Legal Analysis}, 4(2):471--514, 2012.

\bibitem{hadfield2013law}
Gillian~K Hadfield and Barry~R Weingast.
\newblock Law without the state: legal attributes and the coordination of
  decentralized collective punishment.
\newblock {\em Journal of Law and Courts}, 1(1):3--34, 2013.

\bibitem{hadfield2014microfoundations}
Gillian~K Hadfield and Barry~R Weingast.
\newblock Microfoundations of the rule of law.
\newblock {\em Annual Review of Political Science}, 17:21--42, 2014.

\bibitem{hadfield2019legible}
Dylan Hadfield-Menell, McKane Andrus, and Gillian Hadfield.
\newblock Legible normativity for ai alignment: The value of silly rules.
\newblock In {\em Proceedings of the 2019 AAAI/ACM Conference on AI, Ethics,
  and Society}, pages 115--121. ACM, 2019.

\bibitem{henrich2004demography}
Joseph Henrich.
\newblock Demography and cultural evolution: how adaptive cultural processes
  can produce maladaptive losses - the tasmanian case.
\newblock {\em American Antiquity}, 69(2):197--214, 2004.

\bibitem{herrmann2007humans}
Esther Herrmann, Josep Call, Mar{\'\i}a~Victoria Hern{\'a}ndez-Lloreda, Brian
  Hare, and Michael Tomasello.
\newblock Humans have evolved specialized skills of social cognition: The
  cultural intelligence hypothesis.
\newblock {\em science}, 317(5843):1360--1366, 2007.

\bibitem{hughes2018inequity}
Edward Hughes, Joel~Z Leibo, Matthew Phillips, Karl Tuyls, Edgar
  Due{\~n}ez-Guzman, Antonio~Garc{\'\i}a Casta{\~n}eda, Iain Dunning, Tina Zhu,
  Kevin McKee, Raphael Koster, et~al.
\newblock Inequity aversion improves cooperation in intertemporal social
  dilemmas.
\newblock In {\em Advances in neural information processing systems}, pages
  3326--3336, 2018.

\bibitem{jaderberg2016reinforcement}
Max Jaderberg, Volodymyr Mnih, Wojciech~Marian Czarnecki, Tom Schaul, Joel~Z
  Leibo, David Silver, and Koray Kavukcuoglu.
\newblock Reinforcement learning with unsupervised auxiliary tasks.
\newblock {\em arXiv preprint arXiv:1611.05397}, 2016.

\bibitem{kleiman2016coordinate}
Max Kleiman-Weiner, Mark~K Ho, Joseph~L Austerweil, Michael~L Littman, and
  Joshua~B Tenenbaum.
\newblock Coordinate to cooperate or compete: abstract goals and joint
  intentions in social interaction.
\newblock In {\em CogSci}, 2016.

\bibitem{leibo2019autocurricula}
Joel~Z Leibo, Edward Hughes, Marc Lanctot, and Thore Graepel.
\newblock Autocurricula and the emergence of innovation from social
  interaction: A manifesto for multi-agent intelligence research.
\newblock {\em arXiv preprint arXiv:1903.00742}, 2019.

\bibitem{leibo2017multi}
Joel~Z Leibo, Vinicius Zambaldi, Marc Lanctot, Janusz Marecki, and Thore
  Graepel.
\newblock Multi-agent reinforcement learning in sequential social dilemmas.
\newblock In {\em Proceedings of the 16th Conference on Autonomous Agents and
  MultiAgent Systems}, pages 464--473. International Foundation for Autonomous
  Agents and Multiagent Systems, 2017.

\bibitem{lerer2017maintaining}
Adam Lerer and Alexander Peysakhovich.
\newblock Maintaining cooperation in complex social dilemmas using deep
  reinforcement learning.
\newblock {\em arXiv preprint arXiv:1707.01068}, 2017.

\bibitem{Littman94markovgames}
M.~L. Littman.
\newblock Markov games as a framework for multi-agent reinforcement learning.
\newblock In {\em Proceedings of the 11th International Conference on Machine
  Learning (ICML)}, pages 157--163, 1994.

\bibitem{mahmoud2016cooperation}
Samhar Mahmoud, Simon Miles, and Michael Luck.
\newblock Cooperation emergence under resource-constrained peer punishment.
\newblock In {\em Proceedings of the 2016 International Conference on
  Autonomous Agents \& Multiagent Systems}, pages 900--908. International
  Foundation for Autonomous Agents and Multiagent Systems, 2016.

\bibitem{mathew2013small}
S~Mathew, R~Boyd, and M~van Veelen.
\newblock Human cooperation among kin and close associates may require
  enforcement of norms by third parties.
\newblock In Peter~J. Richerson and Morten~H. Christiansen, editors, {\em
  Cultural Evolution}. MIT Press, Cambridge, MA, 2013.

\bibitem{meyer2009food}
Victor~Benno Meyer-Rochow.
\newblock Food taboos: their origins and purposes.
\newblock {\em Journal of Ethnobiology and Ethnomedicine}, 5(18), 2009.

\bibitem{morales2013automated}
Javier Morales, Maite Lopez-Sanchez, Juan~A Rodriguez-Aguilar, Michael
  Wooldridge, and Wamberto Vasconcelos.
\newblock Automated synthesis of normative systems.
\newblock In {\em Proceedings of the 2013 international conference on
  Autonomous agents and multi-agent systems}, pages 483--490. International
  Foundation for Autonomous Agents and Multiagent Systems, 2013.

\bibitem{morris2019norm}
Andreasa Morris-Martin, Marina De~Vos, and Julian Padget.
\newblock Norm emergence in multiagent systems: a viewpoint paper.
\newblock {\em Autonomous Agents and Multi-Agent Systems}, 33(6):706--749,
  2019.

\bibitem{oord2018representation}
Aaron van~den Oord, Yazhe Li, and Oriol Vinyals.
\newblock Representation learning with contrastive predictive coding.
\newblock {\em arXiv preprint arXiv:1807.03748}, 2018.

\bibitem{ostrom1990governing}
Elinor Ostrom.
\newblock {\em Governing the commons: The evolution of institutions for
  collective action}.
\newblock Cambridge university press, 1990.

\bibitem{perolat2017multi}
Julien Perolat, Joel~Z Leibo, Vinicius Zambaldi, Charles Beattie, Karl Tuyls,
  and Thore Graepel.
\newblock A multi-agent reinforcement learning model of common-pool resource
  appropriation.
\newblock In {\em Advances in Neural Information Processing Systems}, pages
  3643--3652, 2017.

\bibitem{peysakhovich2017consequentialist}
Alexander Peysakhovich and Adam Lerer.
\newblock Consequentialist conditional cooperation in social dilemmas with
  imperfect information.
\newblock {\em arXiv preprint arXiv:1710.06975}, 2017.

\bibitem{peysakhovich2018prosocial}
Alexander Peysakhovich and Adam Lerer.
\newblock Prosocial learning agents solve generalized stag hunts better than
  selfish ones.
\newblock In {\em Proceedings of the 17th International Conference on
  Autonomous Agents and MultiAgent Systems}, pages 2043--2044. International
  Foundation for Autonomous Agents and Multiagent Systems, 2018.

\bibitem{powell2009late}
Adam Powell, Stephen Shennan, and Mark~G Thomas.
\newblock Late pleistocene demography and the appearance of modern human
  behavior.
\newblock {\em Science}, 324(5932):1298--1301, 2009.

\bibitem{richerson1998ultrasociality}
Peter Richerson and Robert Boyd.
\newblock The evolution of human ultra-sociality.
\newblock In I.~Eibl-Eibisfeldt and F.~Salter, editors, {\em Ideology, Warfare,
  and Indoctrinability}, pages 71--95. Berghan Books, Oxford, 1998.

\bibitem{riedl2012chimps}
Katrin Riedl, Keith Jensen, Josep Call, and Michael Tomasello.
\newblock No third-party punishment in chimpanzees.
\newblock {\em Proceedings of the National Academy of Sciences of the United
  States of America}, 109(37):14824--14829, 2012.

\bibitem{shapley1953stochastic}
L.~S. Shapley.
\newblock {Stochastic Games}.
\newblock {\em In Proc. of the National Academy of Sciences of the United
  States of America}, 1953.

\bibitem{sutton2018reinforcement}
Richard~S Sutton and Andrew~G Barto.
\newblock {\em Reinforcement learning: An introduction}.
\newblock MIT press, 2018.

\bibitem{tomasello2013review}
Michael Tomasello and Amrisha Vaish.
\newblock Origins of human cooperation and morality.
\newblock {\em Annual Review of Psychology}, 64(1):231--255, 2013.

\bibitem{wang2019evolving}
Jane~X Wang, Edward Hughes, Chrisantha Fernando, Wojciech~M Czarnecki, Edgar~A
  Du{\'e}{\~n}ez-Guzm{\'a}n, and Joel~Z Leibo.
\newblock Evolving intrinsic motivations for altruistic behavior.
\newblock In {\em Proceedings of the 18th International Conference on
  Autonomous Agents and MultiAgent Systems}, pages 683--692. International
  Foundation for Autonomous Agents and Multiagent Systems, 2019.

\bibitem{wiessner2005norm}
Polly Wiessner.
\newblock Norm enforcement among the ju$/'$hoansi bushmen.
\newblock {\em Human Nature}, 16(2):115--145, 2005.

\end{thebibliography}
\bibliographystyle{plain}

\end{document}